\documentclass[pdftex,twocolumn,epjc3]{svjour3}          

\RequirePackage[T1]{fontenc}
\pdfoutput=1

\smartqed  

\RequirePackage{graphicx}
\RequirePackage{mathptmx}      
\RequirePackage{flushend}
\RequirePackage{chemformula}
\RequirePackage{siunitx}
\RequirePackage[super]{nth}
\RequirePackage{lineno}
\RequirePackage[numbers,sort&compress]{natbib}
\RequirePackage[colorlinks,citecolor=blue,urlcolor=blue,linkcolor=blue]{hyperref}

\journalname{Eur. Phys. J. C}
\newcommand{\nT}{XENONnT }
\newcommand{\rbttm}{$R_\mathrm{bot}$}
\newcommand{\vtfc}{$V_\mathrm{top}$}
\newcommand{\mciv}{$M_\mathrm{CIV}$}
\newcommand{\kr}{\ch{^{83m}Kr} }
\newcommand{\ar}{\ch{^{37}Ar} }
\newcommand{\rn}{\ch{^{222}Rn} }
\newcommand{\rb}{\ch{^{83}Rb} }

\newcommand{\figref}[1]{Fig.~\ref{#1}}
\newcommand{\secref}[1]{Sect.~\ref{#1}}

\begin{document}

\title{Design and performance of the field cage for the XENONnT experiment}

\author{E.~Aprile\thanksref{columbia} \and K.~Abe\thanksref{tokyo} \and S.~Ahmed Maouloud\thanksref{paris} \and L.~Althueser\thanksref{munster} \and B.~Andrieu\thanksref{paris} \and E.~Angelino\thanksref{torino} \and J.~R.~Angevaare\thanksref{nikhef} \and V.~C.~Antochi\thanksref{stockholm} \and D.~Ant\'on Martin\thanksref{chicago} \and F.~Arneodo\thanksref{nyuad} \and L.~Baudis\thanksref{zurich} \and A.~L.~Baxter\thanksref{purdue} \and M.~Bazyk\thanksref{subatech} \and L.~Bellagamba\thanksref{bologna} \and R.~Biondi\thanksref{heidelberg} \and A.~Bismark\thanksref{zurich} \and E.~J.~Brookes\thanksref{nikhef} \and A.~Brown\thanksref{freiburg} \and S.~Bruenner\thanksref{nikhef} \and G.~Bruno\thanksref{subatech} \and R.~Budnik\thanksref{wis} \and T.~K.~Bui\thanksref{tokyo} \and C.~Cai\thanksref{tsinghua} \and J.~M.~R.~Cardoso\thanksref{coimbra} \and D.~Cichon\thanksref{heidelberg} \and A.~P.~Cimental~Ch\'avez\thanksref{zurich} \and A.~P.~Colijn\thanksref{nikhef} \and J.~Conrad\thanksref{stockholm} \and J.~J.~Cuenca-Garc\'ia\thanksref{zurich} \and J.~P.~Cussonneau\thanksref{subatech,deceased}
% ...
}

\thankstext{email_ft}{e-mail: francesco.toschi@kit.edu}
\thankstext{email_sl}{e-mail: sebastian.lindemann@physik.uni-freiburg.de}
\thankstext{email_mk}{e-mail: kobayashi.masatoshi@isee.nagoya-u.ac.jp}
\thankstext{deceased}{Deceased}
\thankstext{alsoatroma}{Also at INFN - Roma Tre, 00146 Roma, Italy}
\thankstext{alsoatcoimbrapoli}{Also at Coimbra Polytechnic - ISEC, 3030-199 Coimbra, Portugal}
\thankstext{alsoatuniheidelberg}{Also at Physikalisches Institut, Universit\"at Heidelberg, Heidelberg, Germany}

\institute{Physics Department, Columbia University, New York, NY 10027, USA\label{columbia} \and 
Kamioka Observatory, Institute for Cosmic Ray Research, and Kavli Institute for the Physics and Mathematics of the Universe (WPI), University of Tokyo, Higashi-Mozumi, Kamioka, Hida, Gifu 506-1205, Japan\label{tokyo} \and 
LPNHE, Sorbonne Universit\'{e}, CNRS/IN2P3, 75005 Paris, France\label{paris} \and 
Institut f\"ur Kernphysik, Westf\"alische Wilhelms-Universit\"at M\"unster, 48149 M\"unster, Germany\label{munster} \and 
INAF-Astrophysical Observatory of Torino, Department of Physics, University  of  Torino and  INFN-Torino,  10125  Torino,  Italy\label{torino} \and 
Nikhef and the University of Amsterdam, Science Park, 1098XG Amsterdam, Netherlands\label{nikhef} \and 
Oskar Klein Centre, Department of Physics, Stockholm University, AlbaNova, Stockholm SE-10691, Sweden\label{stockholm} \and 
Department of Physics \& Kavli Institute for Cosmological Physics, University of Chicago, Chicago, IL 60637, USA\label{chicago} \and 
New York University Abu Dhabi - Center for Astro, Particle and Planetary Physics, Abu Dhabi, United Arab Emirates\label{nyuad} \and 
Physik-Institut, University of Z\"urich, 8057  Z\"urich, Switzerland\label{zurich} \and 
Department of Physics and Astronomy, Purdue University, West Lafayette, IN 47907, USA\label{purdue} \and 
SUBATECH, IMT Atlantique, CNRS/IN2P3, Universit\'e de Nantes, Nantes 44307, France\label{subatech} \and 
Department of Physics and Astronomy, University of Bologna and INFN-Bologna, 40126 Bologna, Italy\label{bologna} \and 
Max-Planck-Institut f\"ur Kernphysik, 69117 Heidelberg, Germany\label{heidelberg} \and 
Physikalisches Institut, Universit\"at Freiburg, 79104 Freiburg, Germany\label{freiburg} \and 
Department of Particle Physics and Astrophysics, Weizmann Institute of Science, Rehovot 7610001, Israel\label{wis} \and 
Department of Physics \& Center for High Energy Physics, Tsinghua University, Beijing 100084, China\label{tsinghua} \and 
LIBPhys, Department of Physics, University of Coimbra, 3004-516 Coimbra, Portugal\label{coimbra}
% ...
}

\date{Received: date / Accepted: date}

\maketitle
\sloppy

\begin{abstract}
    The precision in reconstructing events detected in a dual-phase time projection chamber depends on an homogeneous and well understood electric field within the liquid target.
    In the XENONnT TPC the field homogeneity is achieved through a double-array field cage, consisting of two nested arrays of field shaping rings connected by an easily accessible resistor chain.
    Rather than being connected to the gate electrode, the topmost field shaping ring is independently biased, adding a degree of freedom to tune the electric field during operation.
    Two-dimensional finite element simulations were used to optimize the field cage, as well as its operation. 
    Simulation results were compared to \kr calibration data.
    This comparison indicates an accumulation of charge on the panels of the TPC which is constant over time, as no evolution of the reconstructed position distribution of events is observed.
    The simulated electric field was then used to correct the charge signal for the field dependence of the charge yield.
    This correction resolves the inconsistent measurement of the drift electron lifetime when using different calibrations sources and different field cage tuning voltages.
\end{abstract}

\section{Introduction}

The strongest direct constraints on dark matter in the form of weakly interacting massive particles (WIMPs) come from noble liquid-gas dual-phase time projection chambers (TPCs) \cite{XENON:2023sxq, xenon1t_results, lz_results, pandax_results, darkside_results, cevns}.
The XENONnT experiment, located at the INFN Laboratori Nazionali del Gran Sasso (LNGS) in central Italy, deploys a dual-phase TPC with a liquid xenon (LXe) target of \SI{5.9}{\tonne} and set an upper limit on the spin-independent WIMP-nucleon elastic scattering cross section down to \SI{2.58e-47}{\square{cm}} for a \SI{28}{\GeV/\square c} WIMP mass at \SI{90}{\percent} confidence level \cite{XENON:2023sxq}.

A particle interacting in the liquid xenon target produces a prompt scintillation light signal (S1) and frees ionization electrons.
The S1 vacuum-ultraviolet (VUV) photons are detected by a top and a bottom array of photomultiplier tubes (PMTs), while the electrons drift upwards following the electric drift field created by a cathode and a gate electrode.
They are then accelerated into a high electric field region between gate and anode.
There they are extracted into the gaseous phase and produce a secondary proportional scintillation signal (S2) before being collected on the anode electrode.
The localized nature of the S2 signal allows an $\left(x,y\right)$-position reconstruction based on the detected light distribution in the top PMT array, while the time difference between the S1 and the S2 signal gives an estimate for the $z$ coordinate.
The ratio between S1 and S2 provides information about the nature of the underlying interaction.
For a given S1 signal, nuclear recoils (NRs) of WIMP or neutron interactions are characterized by a smaller S2 signal than electronic recoils (ERs) from beta or gamma interactions.

\begin{figure}[!b]
    \centering
    \includegraphics[width=\columnwidth]{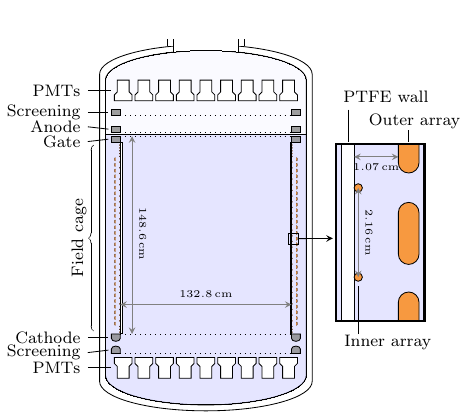}
    \caption{Sketch of the XENONnT TPC. The zoom-in shows a detail of the double-array structure of the field cage, whose implementation is shown in \figref{fig:field_cage}.}
    \label{fig:xenonnt_sketch}
\end{figure}

The electric field at the interaction point in the LXe also affects the signal ratio S2/S1.
For this reason, a homogeneous and well understood electric drift field is crucial for a good discrimination between NR and ER events, and to achieve the best sensitivity for a WIMP search.
The electric fields of the XENONnT TPC are produced by a set of five electrodes (anode, gate, cathode and two screening electrodes) and a field cage enclosing the active volume.
The field cage consists of an inner and an outer array of concentric conductive rings connected by two redundant resistor chains.
A sketch of the TPC with the position of the electrodes and the field cage is shown in \figref{fig:xenonnt_sketch}.

This paper focuses on the simulation and design of the field cage for the XENONnT experiment, with particular emphasis on its improvements with respect to the predecessor experiment, XENON1T \cite{XENON:2017lvq}. 
The design and implementation of the field cage are described in \secref{Section:mech_fc}.
The field simulation setup and the optimization of the resistor chain are summarized in \secref{Section:field_sim}, focusing on the freedom to tune the drift field by changing the voltage applied to the topmost ring of the field cage, treated as an independent electrode.
Finally, the field cage tuning results of XENONnT are discussed in \secref{Section:data_comparison}.
This section also shows the matching of data to simulations of the XENONnT electric field during the first science run, which includes a charge-up component on the TPC reflective walls.

\section{The \nT field cage} \label{Section:mech_fc}
A WIMP scattering in LXe is expected to produce a small scintillation signal S1, hence it is crucial to maximize the light collection efficiency (LCE) of the detector.
In addition to the use of VUV-reflective polytetrafluorethylen (PTFE) walls enclosing the full instrumented target \cite{yamashita_ptfe}, \nT deploys electrodes which are highly transparent to light.
This was achieved using a parallel wire grid design with an optical transparency exceeding \SI{95}{\percent} \cite{nt_instrument_paper}.
The electrodes need to sustain high voltages, as high electric drift fields are known to reduce the fraction of ER events misclassified as NRs, improve the discrimination power between single and multiple scatter, and reduce the maximum electron drift time, limiting the accidental coincidence background \cite{LUX:2020car}.
The design drift field of \nT was \SI{200}{\V\per\cm}, aiming at a larger value than achieved in XENON1T, while considering the past difficulties for dual-phase TPCs in reaching high voltages at the cathode~\cite{Rebel:2014uia}.

The optical transparency of the electrodes translates into a significant field leakage into the drift volume of the extraction field from above the gate and of the reverse field from below the cathode.
This results in an inhomogeneous field within the active volume, leading to a spatially-dependent S2/S1 signal ratio. 
This spatial dependence negatively impacts the discrimination power between signal-like nuclear recoils and background-like electronic recoils, ultimately affecting the final sensitivity to WIMPs \cite{LUX:2020car}.
The field cage plays a crucial role in addressing the problem of field inhomogeneity, forcing a constant voltage gradient within the active volume and effectively mitigating the field leakage through the electrodes.

\begin{figure}[t!]
    \centering
    \includegraphics[width=\columnwidth]{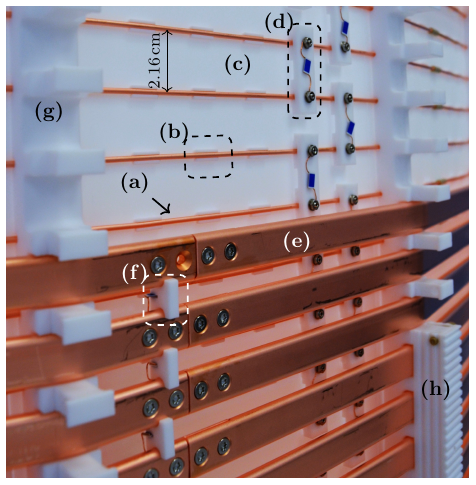}
    \caption{View of the XENONnT field cage from the outside of the TPC during assembly in the clean room. It is possible to discern the different elements: the inner array rings (a) clipped in the notches (b) on the sliding PTFE panels (c) and connected via the resistor chain (d). The outer rings array (e) and its resistor chain (f) are also visible. The pillars (g) are still open as this picture was taken during the assembly. During nominal operation, covers (h) are placed to fix the outer rings. The indicated dimension is given at liquid xenon temperature.}
    \label{fig:field_cage}
\end{figure}

The field cage is composed of an inner and outer set of oxygen-free high conductivity (OFHC, \SI{99.99}{\percent}) copper rings that are connected by a chain of resistors and enclose the entire length of the TPC. 
It is positioned on the outside of the reflecting panels to prevent scintillation photons from being lost due to the photoelectric effect as they hit the copper rings of the field cage, which would reduce the LCE and release single electrons.
A section of the field cage, along with the chain of resistors and various PTFE parts, is shown in \figref{fig:field_cage}.

The double-array structure of the XENONnT field cage introduces a novel approach in the field of dual-phase TPCs.
The rigid outer rings act as structural elements, while the smaller inner rings, which come into contact with the PTFE reflectors, facilitate charge removal. 
The small dimension of the inner rings is necessary to minimize the inevitable local field distortion induced by the presence of conductive elements close to the active volume.
For the same reason, despite their wider surface area, the outer rings have a non-discernible impact on the local drift field.

The \nT field cage was constructed to make contact with the exterior of the PTFE walls wherever feasible.
This decision was prompted by the observation in XENON1T of an inward push in the reconstructed (or observed) radial position of the events correlated to the $\left(x,y\right)$ geometry of the TPC \cite{XENON:2019ykp}. 
The radial distortion was explained as a charge-up of the PTFE walls, exhibiting a time dependence. 
The azimuthal dependence, known as ``bite-structure'' showed a stronger inward push around the panels than around the pillars. 
This effect was attributed to a smaller accumulation of charges on the pillars with respect to the panels, possibly due to a more efficient removal process resulting from the contact of the XENON1T field cage with the pillars. 

The active region of the XENONnT TPC is a prism with a height of \SI{148.6}{\cm} and a 24-sided polygonal cross-section, with \SI{132.8}{\cm} between opposite sides.
%
Two PMT arrays and stacks of electrodes limit the TPC at the top and bottom. 
The electrode stacks as well as the PMT arrays are supported by 24 PTFE pillars.
A total of 24 \SI{3}{\mm} thick PTFE ``sliding'' panels are mounted in between the pillars and they interlock with 24 PTFE ``blocking'' panels directly mounted on the pillars.
A total of 355 clipping notches are incorporated into each of the sliding panels to maintain contact between the inner rings of the field cage and the insulator.
In addition, the sliding panels feature \SI{0.25}{\mm} diameter through-holes at the center of each clipping notch.
They serve the purpose of facilitating the removal of free charges present on the inner side of the wall, as the mobility of electrons along the PTFE surface is expected to be larger than across the material bulk.

The 71 inner rings consist of \SI{2}{\mm} diameter wires, taken from a single OFHC copper spool.
The wire was first stretched around a mock-up of \SI{133.1}{\cm} inner diameter, cut to the right length, and both ends were threaded. 
During installation the ends were connected using polyether ether ketone (PEEK) fasteners, allowing the circumference to be adjusted by a few millimeters. 

The outer field cage array consists of 64 rigid copper rings, each having a 24-sided polygonal shape with a \SI{135.5}{\cm} distance between opposite sides and a cross section of \qtyproduct[product-units=single]{15x5}{\mm} with a \SI{2.5}{\mm} rounding radius.
The outer rings are positioned along $z$ between \num{-7} and \SI{-145}{\cm} ($z=0$ being the vertical position of the gate) and with half pitch offset from the inner field cage rings.
Each outer ring consists of two halves connected by four countersunk M3 stainless steel bolts.
One half ring is meant to be fixed in position and it features two additional holes close to the junction which are used to connect the resistor chains.
The other half ring can be removed for easier access during maintenance.

The geometry of the field cage is mostly constrained by the compact TPC design and its key role for the mechanical stability of the detector.
The minimal radial distance between inner and outer field cage arrays is \SI{8.7}{\mm}. 
The radial position of the inner rings is determined by the PTFE wall they are clipped into, whereas the outer radius of the outer rings is limited by the high-voltage feedthrough (HVFT) running to the cathode along the full length of the TPC.
While a larger radius of the field cage would improve the drift field homogeneity, a smaller distance between outer rings and the grounded stainless steel sleeve of the HVFT increases the risk of discharges.

The vertical distribution of the field cage arrays is constrained at the top and bottom by the position of the gate and cathode frames.
A pitch of \SI{21.6}{\mm} at liquid xenon temperature was chosen to facilitate the assembly of the resistor chain within the reduced intra-array space.
Additional inner rings are included in the design at the top and bottom ends of the field cage: four rings right below the gate and two above the cathode (compare to Figure~\ref{fig:xenonnt_sketch}).
These extra elements are installed with half the normal pitch.
They improve the field homogeneity in regions dominated by edge effects and field leakage coming from the electrodes' transparency.

The voltage divider of the field cage is entirely realized using \SI{5}{\giga\ohm} SMD resistors with \SI{1}{\percent} tolerance by OHMITE \cite{resistor}.
These resistors were already used in XENON1T and extensively tested against failures.
They are arranged in order to ensure a linear potential drop along the $z$-axis.
The inner and outer set of rings have independent resistor chains, which are connected at the top and bottom to form the voltage divider of the field cage.
This minimizes the impact on the electric field in case of a broken resistor, while simplifying its assembly.
Two redundant sets of voltage dividers are implemented on opposite sides of the TPC.

The electrical connection for the outer rings is achieved by clamping the end of a \SI{0.4}{\mm} OFHC wire, soldered to the resistor, to the countersunk holes via an M3 screw.
The resistor is then held in place by a \qtyproduct[product-units=single]{7x7}{\mm} PTFE element inserted between two rings.
Given the small wire diameter of the inner array rings, a spring loaded connection was realized.
Dovetail notches on the reflecting panels support counterpart PTFE pieces on which the resistors are mounted.
This joint establishes a spring-loaded connection using M1.7 set screws in electric contact with both the resistors and the inner copper rings.
The connection was tested for stress due to temperature changes using liquid nitrogen, proving a good reliability.

All the materials used to machine and assemble the field cage and its resistor chains were screened and thoroughly cleaned to ensure radiopurity \cite{nt_screening}.

\section{Electric field simulation} \label{Section:field_sim}

The resistor chain and the design of the field cage was optimized based on the simulation of the electric field of the TPC.
The simulations were performed using COMSOL Multiphysics\textsuperscript{\textregistered} v5.4 \cite{comsol}, in particular the AC/DC module with finite element method (FEM) analysis.
This method involves the discretization of the geometry into smaller elements, an operation known as ``meshing''.
The electrostatic equations are then solved at the vertices, or ``nodes'', of each element and interpolated in between.

Given the great number of simulations needed during the design and optimization of the field cage, as well as the high computational power required for a full 3D simulation, a 2D-axisymmetric model of the detector was implemented.
Decreasing the dimension of the problem reduces the number of nodes needed in order to be able to simulate the full detector. 
However, this excludes non-axisymmetrical features, such as the polygonal structure of the field cage or the wire grid nature of the electrodes.
In the 2D-axisymmetric simulation the TPC is constructed as a cylinder with electrodes made of concentric wires.
While this approximation impacts the electric field close to the electrodes, the expected effect in the detector's active volume was estimated to be marginal by simulating a small-scale TPC using both 3D and 2D-axysimmetric geometry.
Comparing the electric field inside the active volume, a difference lower than \SI{0.5}{\percent} was found.
This is notably smaller than the effect of introducing a charge distribution on the PTFE walls, as discussed in \secref{Section:data_comparison}, and it is thus considered negligible.

The TPC is contained inside a vacuum-insulated double-walled stainless steel cryostat, which acts as a Faraday cage.
This means that the simulation of the TPC environment can be restricted to the grounded inner vessel.
Given the 2D-axisymmetric nature of the simulation, elements that only cover a small azimuthal angle are excluded from the model.
This includes the PTFE pillars, the HVFT to the cathode, and the resistor chains.
These elements were studied separately with local 3D simulations in order to evaluate their impact on the drift field and to assess the risk of breakdown.
The PMTs are approximated by a concentric structure in the 2D simulation.
The impact of this approximation on the drift field is expected to be negligible as they are located behind the screening grids and far from the active volume.

The dimensions of the TPC elements span several orders of magnitude, ranging from the \SI{216}{\um} diameter of the electrode wires up to the \SI{1.5}{\m} length of the reflector panels.
For this reason, the mesh size ranges from \SI{30}{\um} around the electrode wires up to \SI{25}{\mm} in the center of the LXe volume, where the electric field is most uniform.
The final mesh consists of \num{4.8e6}~elements and \num{2.4e6}~nodes.
When the field within the active volume is compared to the same geometry simulated with a coarser mesh, the average difference is within \SI{1}{\percent}, being larger close to the electrodes.
Hence, we conclude that the uncertainty from meshing can be ignored.

As discussed in \secref{Section:mech_fc}, the field cage geometry was strongly constrained by mechanical requirements.
For this reason, the uniformity of the electric drift field was optimized by selecting the voltages applied to the field cage.
If the voltage drop is proportional to the vertical separation of two consecutive field shaping elements, then the voltage gradient is constant and the electric field is uniform.
The voltages applied to the top and bottom of the field cage should match the effective potentials in those positions, which differ from the voltages of gate and cathode due to the field leakage effect previously described.
At the top, this matching is done by independently biasing the topmost inner field cage ring.
This freedom in bias voltage represents an important innovation, as it enables the tuning of the field homogeneity during operation of the filled detector.
This permits adjusting to different electrode configurations or exploring the effect of the field homogeneity on the signal, as done in \secref{Section:data_comparison}.
An additional HVFT would have been necessary for a similar solution at the bottom of the field cage, considering the requirement for voltages as low as \SI{-30}{\kV}.
Such a solution was not implemented.
Instead, a fixed resistance between the bottommost inner field cage array element and the cathode was installed.
As it is not possible to change this resistance once the detector is assembled, its value was optimized considering the possibility that the design cathode voltage might not be reached.
The electric field inside the \nT TPC was simulated using the electrodes' design potentials of \SI{-1}{\kV} at the gate, \SI{6.5}{\kV} at the anode and \SI{-30}{\kV} at the cathode.
Different combinations of the topmost inner field cage ring voltage {\vtfc } and the bottom resistance {\rbttm } were considered.
Fields were simulated for a voltage {\vtfc } between \num{-1.2} and \SI{-0.5}{\kV}, and a resistance {\rbttm } between \num{5} and \SI{10}{\giga\ohm}.

\begin{figure}[t]
    \centering
    \includegraphics[width=\columnwidth]{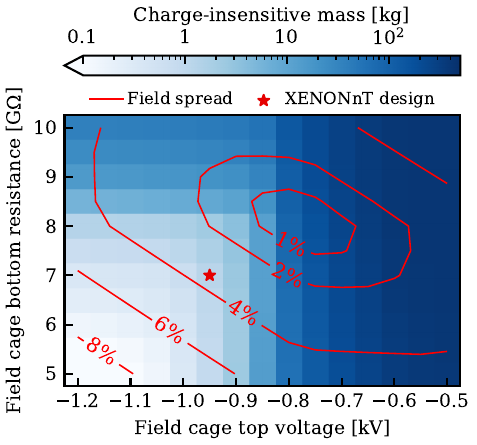}
    \caption{Charge-insensitive mass {\mciv } as a function of the voltage of the topmost inner field cage ring {\vtfc } and the resistance {\rbttm } between the bottom of the field cage and the cathode. The contour lines represent the relative drift field spread within the \SI{4}{\tonne} fiducial volume. The red star shows the configuration picked for XENONnT, with \vtfc$=$\SI{-0.95}{\kV} and \rbttm$=$\SI{7}{\giga\ohm}.}
    \label{fig:field_spread}
\end{figure}

Two independent figures of merit were used in order to evaluate the performance of the different configurations: the field spread within the \SI{4}{\tonne} fiducial volume (FV) as defined in \cite{nt_mc}, and the size of the charge-insensitive volume (CIV).
The field spread is defined as the difference between the \nth{5} and \nth{95} percentile of the electric field magnitude divided by its mean.
The charge-insensitive volume is a region of the detector characterized by the complete or partial loss of the ionization electrons.
The electrons freed in such a volume follow the electric field lines ending on the PTFE walls, accumulating on the wall and thus not producing S2 signals.
The reverse field region between the cathode and the bottom PMT array is an example of irreducible CIV, and it is therefore ignored in this discussion.
The CIV is calculated by propagating the electrons along the simulated field lines from different positions within the TPC active volume and checking whether they end up on the wall surface or reach the liquid-gas interface.
These figures of merit were computed using the custom module \texttt{PyCOMes} \cite{pycomes}, developed to handle COMSOL output format and perform fast calculations of field lines and electron propagation.
The mass of liquid xenon inside the CIV (\mciv) is shown in the (\vtfc, \rbttm) parameter space in \figref{fig:field_spread}.
When the topmost inner field cage ring is biased more positively than the gate electrode, the electrons drifting through the TPC are more strongly attracted to it. 
This improves the uniformity of the drift field within a limited range of voltages. As the bias voltage increases further relative to the gate electrode, the field lines begin to terminate on the PTFE wall, and at approximately $V_\mathrm{top}=-0.85\,\mathrm{kV}$, the CIV abruptly increases.
A larger CIV is also observed for high values of \rbttm.
This is due to an increasing local field distortion in the bottom edge of the detector.

Compromising between field homogeneity and CIV, the values $V_\mathrm{top}=-0.95\,\mathrm{kV}$ and $R_\mathrm{bot}=7\,\mathrm{G\Omega}$ were chosen for the \nT design field.
This corresponds to $M_\mathrm{CIV}=1.2\,\si{\kg}$ and a field spread of \SI{3.5}{\percent}.
We checked the performance of the electric field with the bottom resistance value {\rbttm } for different configurations of the electrode voltages, with the result that the chosen resistance performs sufficiently well for a wide range of scenarios.

\section{Comparison to data} \label{Section:data_comparison}
The \nT detector is periodically calibrated using \ch{^{83m}Kr}.
The metastable isotope has a half-life of \SI{1.83}{\hour} and decays via a two-step transition of \SI{32.2}{\keV} and \SI{9.4}{\keV} with an intervening half-life of \SI{157}{\ns} \cite{kr83m}.
This source is used to monitor the spatial response of the detector and its time evolution, assuming its homogeneous distribution \cite{kr_lux,XENON:2019ykp}.
It is therefore possible to compare the observed \kr event distribution to the expected one from simulations.
The simulated distribution comes from a set of electrons uniformly produced within the active volume.
Each electron is propagated according to the electric field map including diffusion and drift values as coming from literature \cite{exo_diffusion,nikhef_diffusion}.
The $\left(x,y\right)$-position is the electron location at the liquid-gas interface including the position reconstruction resolution, while the $z$ information is derived from the drift time.

\subsection{SR0 field and wall charge-up matching}
During the commissioning phase of the experiment, a short circuit occurred between the cathode and the bottom screen electrode, limiting the voltage of the cathode.
For the first science run (SR0), the electrodes were set to a voltage of \SI{0.3}{\kV} at the gate, \SI{4.9}{\kV} at the anode and \SI{-2.75}{\kV} at the cathode.
This resulted in an average electric drift field of $~$\SI{23}{\V\per\cm}.
The topmost inner field cage ring voltage was set to $V_\mathrm{top}=0.65\,\mathrm{kV}$, which was optimized based on simulations, by means of the procedure described in \secref{Section:field_sim}.

\begin{figure}[b]
    \includegraphics[width=\columnwidth]{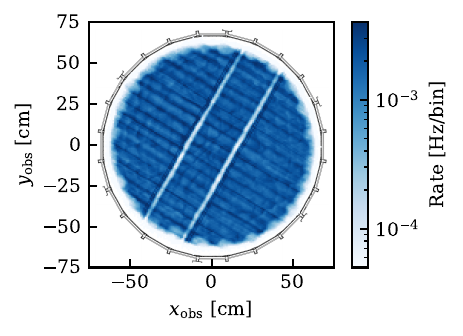}
    \caption{Reconstructed $\left(x,y\right)$-position distribution of \kr events. The distortion at high radii follows the distribution of PTFE pillars and panels, the cross section of which are overlaid in the figure. The diagonal features crossing the TPC result from the transverse wires of the gate electrode and the distribution of the PMTs in the top array.}
    \label{fig:bite_structure}
\end{figure}

\begin{figure}[!t]
    \centering  \includegraphics[width=0.97\columnwidth]{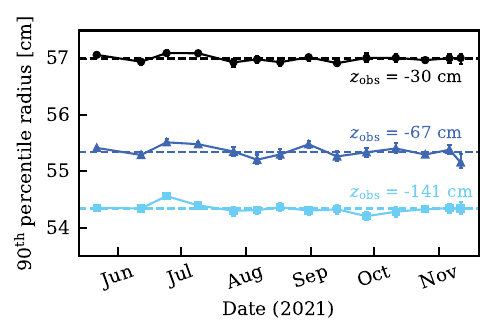}
    \caption{Evolution of the \nth{90} percentile of the radial distribution at three different $z$ bins centered around the indicated value and with a bin width of \SI{7}{\cm}.}
    \label{fig:time_dependence}
\end{figure}

Similarly to XENON1T, an azimuthally dependent distortion at high radii is observed in the \kr distribution, reflecting the 24-sided-polygonal structure of the PTFE walls.
Nevertheless, the distortion shows a different behaviour than what was observed by XENON1T.
In \nT the events around the pillars are pushed more inwards than the events around the panels, leading to a localized strong reduction of the rate, as shown in \figref{fig:bite_structure}.
The observed ``bite-structure'' supports the XENON1T hypothesis of PTFE charge-up discussed in \secref{Section:mech_fc}, which drove the field cage design of XENONnT.
While the field cage rings touch both panels and pillars, the panels are expected to release accumulated charges more easily than pillars because of their thinner geometry and the presence of through-holes. 
The efficient charge removal is supported by the absence of time-dependent features in the reconstructed radial position of the \kr events.
Figure~\ref{fig:time_dependence} shows the evolution over SR0 of the \nth{90} percentile of the radius distribution of \kr events in three slices of $z$.
Unlike previous experiments \cite{lux_3d_field,XENON:2019ykp}, no increase of the inward push is observed over time.
In addition, the observed $\left(x,y\right)$-distribution has two symmetric features crossing the TPC as a result of the transverse wires of the gate locally deflecting the electrons when drifting towards the liquid-gas interface.
These wires were installed both at the gate and anode electrodes to counteract wire deformation under electrostatic force \cite{nt_instrument_paper}.
The regular pattern perpendicular to these features is due to a combination of the wire grid of the anode electrode with the geometrical configuration of the PMTs in the top array.

\begin{figure}[b]
    \centering 
    \includegraphics[width=0.97\columnwidth]{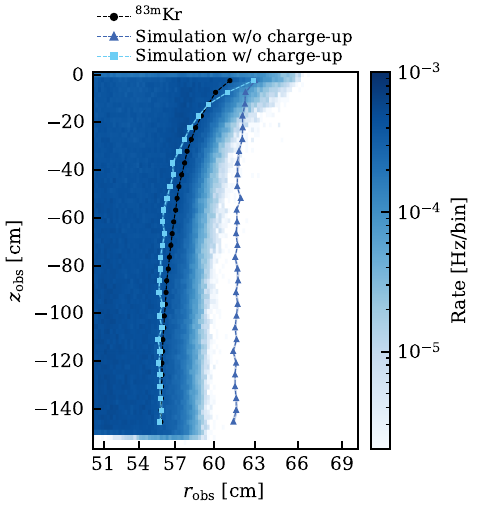}
    \caption{($r$,$z$) distribution of \kr events near the walls of the TPC. The \nth{90} percentile of the radial distribution along 30 bins of $z$ is shown in black. The same quantities coming from the simulation with and without PTFE reflector charge-up are shown in blue and orange, respectively.}
    \label{fig:match_field}
\end{figure}

The \nth{90} percentile of the radial distribution $r_{90}$ is evaluated for both \kr data and simulation in 30 bins of $z$ by averaging over the azimuthal angle.
The simulation consists in the propagation of $10^5$ electrons with initial position uniformly distributed in the active volume.
The \nth{90} percentile is sufficiently large not to be affected by the transverse wires, but not sensitive to possible outliers at high percentiles.
A mismatch between data and simulation can be clearly seen in \figref{fig:match_field}, with the difference between $r_{90}$ from data (black circles) and from simulation (blue triangles) being on average \SI{4.7}{\cm}.
The mismatch can be effectively resolved by considering a charge accumulation on the PTFE walls, as already demonstrated in previous works \cite{lux_3d_field}.
The corresponding surface charge density $\sigma_\mathrm{w}$ is determined by matching the observed radial distribution with simulations including a charge distribution at the walls.
This density is parameterized using a linear function:
\begin{linenomath}
\begin{equation}
\label{eq:linear_charge}
\sigma_\mathrm{w}=\lambda \cdot \frac{\left|z\right|}{h_\mathrm{TPC}}+\sigma_\mathrm{top},
\end{equation}
\end{linenomath}
where $h_\mathrm{TPC} = 148.6\,\mathrm{cm}$ is the height of the TPC, $\sigma_\mathrm{top}$ is the surface charge density at the top of the panels and $\lambda$ is the charge density difference between the top and bottom of the panels, i.e., $\sigma_\mathrm{bot} = \sigma_\mathrm{top} + \lambda$. 
A linear model describes to first order the observations reported in \cite{lux_3d_field}.
Field and electron-propagation simulations were performed for $\sigma_\mathrm{top}$ between \num{-2} and \SI{1.5}{\micro\coulomb\per\meter\squared} and for $\lambda$ between \num{-1} and \SI{1.5}{\micro\coulomb\per\meter\squared}, both with steps of \SI{0.1}{\micro\coulomb\per\meter\squared}.
For each combination of $\sigma_\mathrm{top}$ and $\lambda$, $r_{90}^\mathrm{sim}$ was calculated for $n_z=30$ bins in $z$ along the TPC as described above.
The chi-square was estimated for each simulation in the following way:
\begin{linenomath}
\begin{equation}
    \chi^2=\sum_{i=0}^{n_z-1}\frac{\left(r_{90, i}^\mathrm{obs} - r_{90, i}^\mathrm{sim}\right)^2}{\sigma_{90, i}^2},
\end{equation}
\end{linenomath}
where $\sigma_{90}^2$ is the squared sum of the statistical percentile uncertainties of data and simulations.
The $\chi^2$ best fit yields 
\begin{linenomath}
\begin{align*}
\sigma_\mathrm{top}&=\left(\num{-0.50(0.06)}\text{\scriptsize{(syst)}}\pm0.02\text{\scriptsize{(stat)}}\right)\,\si{\micro\coulomb\per\meter\squared},\\
\lambda&=\left(\num{0.40(0.15)}\text{\scriptsize{(syst)}}^{+\num{0.20}}_{\num{-0.10}}\text{\scriptsize{(stat)}}\right)\,\si{\micro\coulomb\per\meter\squared}.
\end{align*}
\end{linenomath}
These values correspond to a surface charge density of \SI{-0.5}{\micro\coulomb\per\meter\squared} at the top of the panels and \SI{-0.1}{\micro\coulomb\per\meter\squared} at the bottom.
The statistical uncertainty was determined by resampling the simulated position distributions for each parameter combination, a technique known as ``bootstrapping'', and then assessing their $\chi^2$ best fit.
The systematic uncertainty was obtained by repeating the $\chi^2$ minimization with different binning in $z$ and percentile values, and taking into account the coarse binning for the $\sigma_\mathrm{top}$ and $\lambda$ parameters. 

\begin{figure}[!t]
    \centering
    \includegraphics[width=0.97\columnwidth]{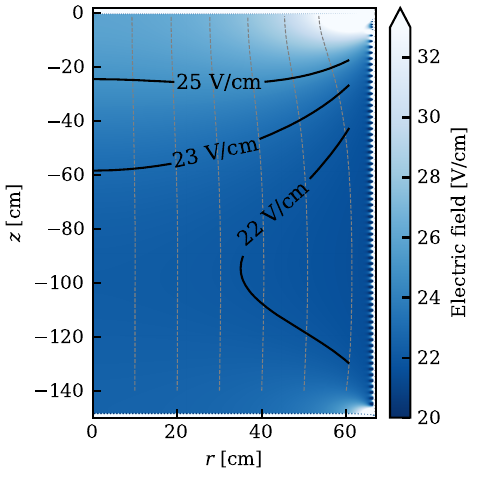}
    \caption{Electric field map determined from 2D-axisymmetric simulations including a linear charge distribution on the PTFE reflectors matched to the radial distribution of \kr events. The black lines indicate the contour of the electric field, while the dashed grey lines are field lines starting at different radii and same $z$.}
    \label{fig:field_map_charged}
\end{figure}

The simulated radial distribution after adding the wall charge-up component is shown as cyan squares in \figref{fig:match_field}, with a maximum difference with respect to the observed distribution of \SI{1.6}{\cm} at the very top of the TPC and \SI{0.3}{\cm} on average.
Including the surface charge density, the predicted field spread is \SI{13.2}{\percent} within the FV and the charge-insensitive mass is \SI{112}{\kg}.
The corresponding SR0 electric field map including charge-up is shown in \figref{fig:field_map_charged}.
The hypothesis that a failure of the resistor chain causes the mismatch between the simulation and the data can be ruled out, as the total resistance of the field cage was measured to be \SI{92(11)}{\giga\ohm}, in good agreement with the expected value of \SI{87.25(5)}{\giga\ohm}.
Moreover, the simulation of the failure of a single resistor shows an insufficient impact on the observed position distribution.

An independent validation of the field map including wall charge-up comes from the measurement of the electron lifetime during SR0. 
The electron lifetime $\tau_{e^-}$ is the characteristic time constant of the exponential decrease of the S2 signal as a function of drift time $t_d$. 
This is due to electrons being trapped by impurities in the liquid xenon.
To determine the electron lifetime $\tau_{e^-}$, an exponential function is fitted to the median of the S2 area across different drift times.
Previous analyses of XENON1T data revealed a discrepancy in the measurement of the electron lifetime when using different radioactive isotopes, such as \kr and \rn \cite{XENON:2019ykp}.
These isotopes vary in their decay products and energy, resulting in different ionization densities within xenon.
For this reason, the electric field affects the charge signal of each calibration source differently, leading to a different spatial dependence of the S2 signal in presence of an inhomogeneous drift field.
The electron lifetime measured according to the above approach is thus an effective value $\tau_{e^-}^\mathrm{eff}$ that includes a relative field effect on the charge yield, $\mathit{Q_\mathrm{y}^\mathrm{rel}}(x,y,z)$:
\begin{figure*}[!ht]
    \centering
    \includegraphics{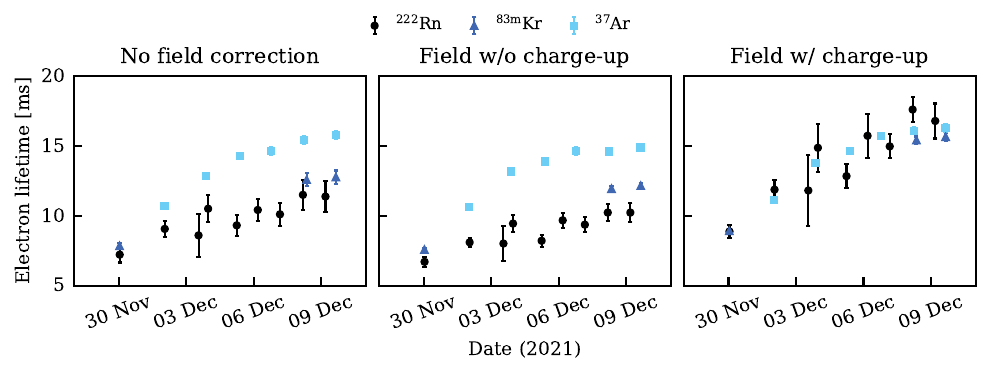}
    \caption{Electron lifetime measured using three different radioactive isotopes: \ar (light blue squares), \kr (blue triangles) and \rn (black circles). The results on the left plot use the uncorrected charge signals, while middle and right plots include a drift field correction based on the field map without and with charge on the PTFE wall, respectively. The data were taken during a simultaneous \ar and \kr calibration after the end of SR0. Due to emanation, minute levels of \rn are present in the detector at all times \cite{nt_screening}.
    }
    \label{fig:elifetime_match}
\end{figure*}
\begin{linenomath}
\begin{equation}
\label{eq:electron_lifetime}
\begin{aligned}
        \mathrm{S2}\left(t_d\right) &=\mathrm{S2}\left(0\right)\cdot\exp{\left(-t_d/\tau_{e^-}^\mathrm{eff}\right)}\\
        &=\mathrm{S2}\left(0\right)\cdot \mathit{Q_\mathrm{y}^\mathrm{rel}}\left(x,y,z\right)\cdot\exp{\left(-t_d/\tau_{e^-}\right)}.
\end{aligned}
\end{equation}
\end{linenomath}
Figure~\ref{fig:elifetime_match} shows the results from the measurement of the electron lifetime during a joint calibration using \ar and \kr sources, which was carried out after the end of SR0 \cite{lower_nt}.
The ``uncorrected'' electron lifetime comes from the exponential fit of the S2 median area, and it corresponds to  $\tau_{e^-}^\mathrm{eff}$ of Eq.~\eqref{eq:electron_lifetime}.
The ``corrected'' lifetime is obtained by weighting the measured S2 area for the relative charge yield $\mathit{Q_\mathrm{y}^\mathrm{rel}}$ as coming from the field map.
This is the best estimate of the true electron lifetime $\tau_{e^-}$.
Since each calibration source is affected differently by the electric field, $\mathit{Q_\mathrm{y}^\mathrm{rel}}$ is estimated for each isotope.
The \kr charge yield is modeled using data from \cite{heidelberg_hexe}.
\ar data are extrapolated to lower electric fields using the results from \cite{pixey}. 
The charge yield of \rn alphas finally is modeled using NEST v1 \cite{nest_paper,nestv1}, which is more consistent with recent measurements than the latest version \cite{heidelberg_hexe}.
When not corrected, the electron lifetimes from different sources (left panel) do not agree among themselves, as this corresponds to assuming a perfectly homogeneous electric drift field.
The lifetimes are corrected using the simulated electric field maps both with (right panel) and without (middle panel) the inclusion of the charge-up of the PTFE walls.
The different measurements are in agreement when charge accumulation is assumed, but when it is not included, the discrepancy is even more pronounced than in the uncorrected case.

\subsection{Impact of the field cage tuning on the drift field}
\begin{figure*}[!t]
    \centering
    \includegraphics[width=\textwidth]{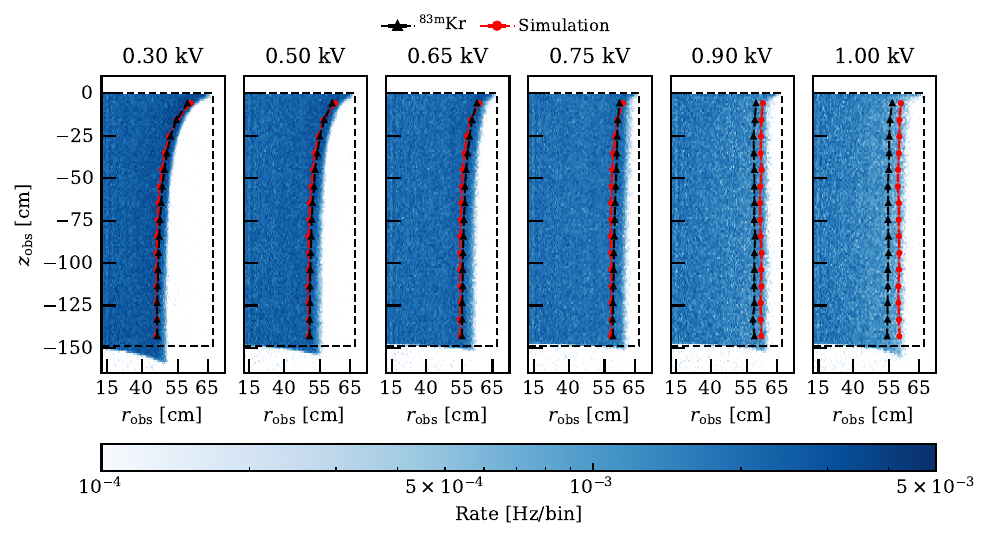}
    \caption{Reconstructed position distribution of \kr events for different voltages of the topmost inner ring of the field cage. 
    Red circles and black triangles are the \nth{90} percentile radial distribution coming from simulation and data, respectively.
    The TPC active volume boundaries are shown as black dashed lines.}
    \label{fig:data_comparison}
\end{figure*}

Dedicated datasets were taken after the end of SR0 to assess the impact of different voltages of the topmost inner field cage ring {\vtfc } on the drift field.
This was varied from \SI{0.3}{\kV} to \SI{1}{\kV} during a \kr calibration, while keeping the voltages of all other electrodes at their SR0 values.
While the electrode voltages mostly impact the magnitude of the electric field, the independent biasing of the topmost inner field cage ring influences primarily its homogeneity.
From simulations, changing {\vtfc } from \SI{0.3}{\kV} (same as gate) to \SI{1}{\kV} translates into a \SI{7}{\percent} stronger field within the FV, while reducing the field spread by \SI{400}{\percent}.
For this reason, this is the first direct measurement of the effect of the field homogeneity on signal production and the transport of S2 electrons in a multi-tonne LXe TPC.

The reconstructed \kr $\left(r,z\right)$ distribution and the \nth{90} percentile radial distribution are shown in \figref{fig:data_comparison} for different field cage tuning voltages.
As {\vtfc } increases, a decreasing radial inward push is observed.
As discussed in \secref{Section:field_sim}, a more positive voltage at the top of the TPC attracts electrons counteracting the inward push, resulting in a more uniform distribution.
However, by increasing {\vtfc } the charge-insensitive volume increases.
The CIV cannot be inferred from the observed position distribution even for $V_\mathrm{top}>0.75\,\mathrm{kV}$, when $>10\,\si{\percent}$ of the total TPC volume is charge-insensitive.
For these configurations, the edge of the position distribution is flat over $z$, showing no inward feature.

The comparison of $r_{90}^\mathrm{obs}$ for different {\vtfc } with the corresponding $r_{90}^\mathrm{sim}$ including the SR0 wall charge distribution returns a good match for voltages \vtfc below or equal to \SI{0.75}{\kV}.
A difference of up to \SI{5}{\cm} in the \nth{90} percentile radial distribution is observed for a {\vtfc } of \SI{0.9}{\kV} and \SI{1}{\kV}.
This hints towards a mismodeling of the charge distribution or the possibility that the charge distribution reaches a new equilibrium for high voltages \vtfc.

\begin{figure}[ht]
    \centering
    \includegraphics[width=0.97\columnwidth]{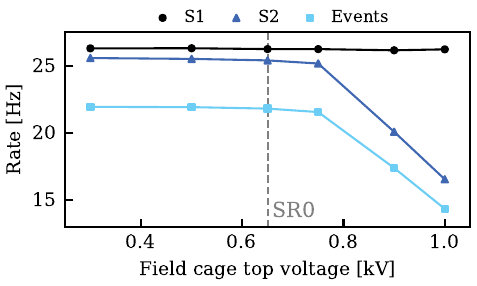}
    \caption{Rate of events (light blue squares), signals S1 (black circles) and S2 signals (blue triangles) for the same \kr source, but different topmost inner field cage ring voltages.
    All rates are corrected for the \rb source decay.
    }
    \label{fig:rate_s1s2}
\end{figure}

The change of the CIV is reflected by the change of the observed event and signal rates for different {\vtfc } values, as shown in \figref{fig:rate_s1s2}.
An event is defined by the pairing of an S1 and S2 signal \cite{lower_nt} passing a loose box \ch{^{83m}Kr} selection on their area, while the individual signals (S1 or S2) are selected within the corresponding range, but without pairing requirement.
The time elapsed between the beginning and the end of the test is around one day. 
Because of the \rb half-life of \SI{86.2}{\day}, a daily decrease in rate of \SI{1.01(12)}{\percent} is considered in the calculation, verified by comparing the rate before and after the test using the same SR0 field configuration.
As expected, the S1 rate is constant for different \vtfc, while S2 and event rates are fairly constant up to \SI{0.75}{\kV}, but quickly drop for larger values.
This observations proves that a fraction of the active volume is charge-insensitive and that this depends on the electric field configuration.
The fast increase of {\mciv } is explained by the anodic behaviour of the topmost inner field cage ring.
At these voltages the drifting electrons are collected on the PTFE walls at the very top of the TPC.
In this situation even a small change of {\vtfc } leads to a large fraction of field lines being lost at the edges, although the impact on the intensity of the field is negligible.

\begin{figure}[t]
    \centering
    \includegraphics[width=\columnwidth]{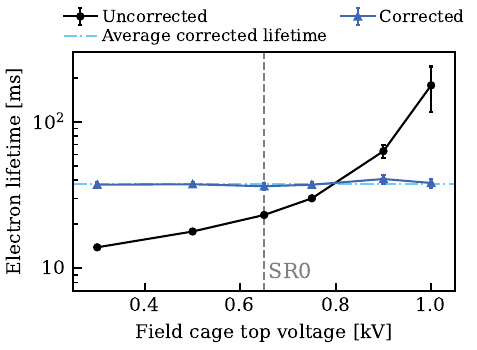}
    \caption{Electron lifetime measured using \kr for different topmost inner field cage ring voltages.
    Blue triangles and black circles are electron lifetime with and without field correction, respectively.
    The cyan dashed line corresponds to \SI{38}{\ms}, which is the average value of measurements with field correction. 
    }
    \label{fig:elife_kr_test}
\end{figure}

Thanks to the large rate of \kr events collected during the test, it is possible to measure the electron lifetime individually for each voltage.
A clear dependence on {\vtfc } is shown in \figref{fig:elife_kr_test}.
The observed increase of the uncorrected electron lifetime (black circles) is explained by the lower electric field at the top of the TPC as {\vtfc } increments, as it is suggested by the field simulations.
A smaller electric field leads to a reduced charge yield, finally resulting in an higher uncorrected electron lifetime.
As these data have been taken within few hours, the fast change of the electron lifetime as {\vtfc } increases cannot be due to a change of the impurity concentration in the liquid xenon.
Similarly, the small variation of the electric drift field for different {\vtfc } values cannot account for the change of an order of magnitude in the electron lifetime \cite{attachment}.
Finally, the uncorrected electron lifetime measured right before and after the test in standard field conditions agrees well, excluding a possible evolution over time of the lifetime.

The effects due to the non-uniform electric field on the uncorrected electron lifetime can be accounted for by simulating the electric field, similarly to what is done in \figref{fig:elifetime_match}.
Figure~\ref{fig:elife_kr_test} shows the corrected electron lifetime for different values of \vtfc. 
The corrected values agree with the constant average of \SI{38}{\ms}, considered to be the true electron lifetime $\tau_{e^-}$ and further demonstrating the capability to correct for the electric field effect solely based on simulations.

\section{Summary}

This work demonstrated a good understanding and effective control of the electric field inside the active volume of the XENONnT TPC.
The novel double-array structure of the field cage allows for mechanical stability, while ensuring contact between the conducting field shaping elements and the PTFE walls, facilitating the removal of charges accumulating over time.
The absence of a time evolution in the distribution of the observed event position confirmed an efficient removal.

The innovative independent voltage bias of the topmost field cage ring makes it possible to match it to the local effective potential, a combination between gate and anode voltages due to field leakage through the gate.
The detector was simulated using the FEM software COMSOL Multiphysics\textsuperscript{\textregistered}, using an approximate 2D-axisymmetric geometry.
The bias voltage of the topmost inner field cage ring and the value of the resistor between the field cage and the cathode were chosen to optimize the charge-insensitive volume and the field homogeneity.

During SR0, the spatial distribution of \kr calibration data was compared to the one calculated based on the electric field simulation.
A linear surface charge density along the PTFE walls of the TPC was included in the field simulation to improve the agreement of the reconstructed position distribution between simulation and data.
The best match to data was obtained with a charge density distribution ranging from \SI{-0.5}{\micro\coulomb\per\meter\squared} at the top of the walls to \SI{-0.1}{\micro\coulomb\per\meter\squared} at the bottom, reducing the average difference between simulated and observed \nth{90} percentile radial distribution from \SI{4.7}{\cm} down to \SI{0.3}{\cm}.
The resulting field map was used to correct the relative charge yield of S2 signals used for the estimation of the electron lifetime from different sources.
This resolved a long-standing discrepancy and further validated the simulations.

A dedicated test to investigate the impact of the topmost inner field cage ring voltage on the field uniformity was performed using the data from a \kr calibration source.
An average difference \SI{<1}{\cm} of the \nth{90} percentile radial distribution is observed between data and simulations when including the reflector charge-up for voltages below \SI{0.75}{\kV}.
Above this value, a deteriorating agreement in the position distribution, together with a strong decrease in event and S2 rates, indicating a significant increase of charge-insensitive volume.
The measured electron lifetime as a function of the topmost inner field cage ring voltage showed an apparent increase of an order of magnitude, which cannot be explained by the change of impurity concentrations.
However, as the S2 signals are corrected for the field dependent charge yields evaluated using the proper electric field map, the electron lifetime measurements for the different runs agree within the uncertainties.

The presented design of the field cage for the XENONnT TPC represents a novelty for the dual-phase TPC technology, allowing for control over the homogeneity of the field while minimizing known effects of charge accumulation on the detector walls.
Together with the good understanding of the electric drift field, this elevates the capability of TPC detectors for dark matter searches improving the sensitivity to WIMPs and potentially setting a new standard in the field.

\begin{acknowledgements}
We gratefully acknowledge support from the National Science Foundation, Swiss National Science Foundation, German Ministry for Education and Research, Max Planck Gesellschaft, Deutsche Forschungsgemeinschaft, Helmholtz Association, Dutch Research Council (NWO), Weizmann Institute of Science, Israeli Science Foundation, Binational Science Foundation, R\'egion des Pays de la Loire, Knut and Alice Wallenberg Foundation, Kavli Foundation, JSPS Kakenhi and JST FOREST Program in Japan, Tsinghua University Initiative Scientific Research Program and Istituto Nazionale di Fisica Nucleare.
This project has received funding/support from the European Union’s Horizon 2020 research and innovation programme under the Marie Sk\l{}odowska-Curie grant agreement No 860881-HIDDeN.
Data processing is performed using infrastructures from the Open Science Grid, the European Grid Initiative and the Dutch national e-infrastructure with the support of SURF Cooperative.
We are grateful to Laboratori Nazionali del Gran Sasso for hosting and supporting the XENON project.
\end{acknowledgements}

\end{document}